\documentstyle[aps,preprint,epsfig]{revtex}

\begin{document}
\preprint{TU-587}
\title{Low-Scale Anomalous U(1) and Decoupling Solution to Supersymmetric
Flavor Problem}
\author{Shinji Komine\thanks{e-mail: komine@tuhep.phys.tohoku.ac.jp},
 Youichi Yamada\thanks{e-mail: yamada@tuhep.phys.tohoku.ac.jp},  and 
Masahiro Yamaguchi\thanks{e-mail: yama@tuhep.phys.tohoku.ac.jp}}
\address{Department of Physics, Tohoku University,
Sendai 980-8578, Japan}
\date{Feburary 2000}
\maketitle
\begin{abstract}
Supersymmetric standard models where the ultraviolet cut-off scale is only a 
few orders of magnitude higher than the electroweak scale are
considered. Phenomenological consequences of this class of models are expected
to be very different from, for example, the conventional supergravity 
scenario.  We apply this idea to a model with an anomalous U(1) gauge group and
construct a viable model in which some difficulties of the
decoupling solution to the supersymmetric flavor problem are ameliorated.
\end{abstract} 

\clearpage


What is the fundamental scale of the unified theory of particle
physics? Usually it is supposed to be near the Planck scale. In fact
this view is supported by the weakly coupled heterotic string theory
where the string scale must be close to the Planck scale
\cite{Ginsparg}. It has recently been recognized, however, that the
fundamental scale can be much lower than the Planck scale, even as low
as around 1 TeV \cite{ADD}.  In this case, the largeness of the Planck
scale is accounted for by extra dimensions with large volume, in which
gravity propagates, while the standard model particles have to be
confined on four dimensional subspace (3-brane). Remarkably this
configuration survives various phenomenological constraints
\cite{ADD,collider,astro,cosmo}.  Furthermore it can be realized in,
for example, type I (or more precisely type I') superstring theory
where the string scale may not directly be related to the Planck scale
\cite{AADD} (See also Ref. \cite{previous} for earlier attempts).
Even when the fundamental scale itself is close to the Planck scale,
special geometrical configuration of extra dimension(s) such as an AdS
slice \cite{RS} may enable us to obtain a model with effectively very
low cut-off scale on a visible brane where the standard model sector
is confined.

In this paper, we would like to consider the situation where the
fundamental scale, or the ultraviolet cut-off scale, 
is much lower than the Planck scale but still lies a
few order of magnitude above the electroweak scale so that low energy
supersymmetry is needed to protect the electroweak scale from
radiative corrections. A supersymmetric standard model with such a
low-scale cut-off will naturally fall into a class of low-scale
supersymmetry breaking models.  An immediate consequence of this class
of models is that the gravitino, the superpartner of the graviton, is
much lighter than other superparticles, and thus tends to be the
lightest superparticle. Moreover soft supersymmetry breaking masses
are given at the low energy scale so that the mass spectrum is in
general quite different from that of high-scale supersymmetry breaking
models.

Inspired by the arguments given above, here we would like to discuss
phenomenological implications of supersymmetric standard models with
such a low fundamental scale. Specifically we consider a model with
anomalous $U(1)_X$ gauge symmetry \cite{DP,BD,MR}. 
It is well-known that appropriate
assignment of the $U(1)_X$ charges provides the decoupling solution of
the supersymmetric flavor problem \cite{CKN}. Namely the first two
generations of squarks and sleptons are assumed to be heavy, thus 
suppress flavor changing neutral
current (FCNC) processes from superparticle loops, while squarks 
and sleptons in the third
generation are set to be at the electroweak scale, {\it i.e.} a few hundred 
GeV: otherwise the large Yukawa couplings for the
third generation would generate large radiative
corrections to the Higgs mass and then we would lose the very motivation
to introduce low-energy 
supersymmetry. The decoupling solution is simple and 
attractive, in particular when symmetries relate the smallness of the
masses of the quarks and leptons in the first two generations with the 
largeness of the masses of the squarks and sleptons in the first
two generations.

 However it has been pointed out that there are several
difficulties in this scenario. First, although the squarks and
sleptons in the first two generations do not influence the running of
the Higgs mass at one-loop level if one ignores small Yukawa
couplings, they do at two-loop level. Thus they cannot be arbitrarily
heavy: rather their masses are severely limited by the naturalness
argument.  In fact, this issue was discussed in detail in Ref.~\cite{DG}
which gave an upper bound of 5 TeV from the condition that the fine
tuning should be less than $10\%$. The second problem is that the
heavy squarks would give negative contribution to the mass squared of
the third generation squarks at two-loop level, driving the mass
squared negative to cause color breaking \cite{AM,AG}. 
It turns out that the bound
is very severe. In fact the masses of the sfermions in the first
two generations must be much smaller than 10 TeV and thus they are
not large enough to solve the FCNC problem, as far as the
mass of the stop is lighter than about 1 TeV. 
 Finally with a mass
spectrum typical in the decoupling solution, the relic abundance of
the lightest superparticle which is assumed to be bino-like neutralino
tends to overclose the Universe \cite{GRR}.  

What we will do in this paper is first to construct a viable model
with the low fundamental scale and then to show that the problems
above are ameliorated in such a framework. Here we should note that
the issue of the color instability in a scenario where
supersymmetry breaking is mediated at a substantially low energy 
was discussed in
Ref.~\cite{AG}, but without an explicit model.  Also a different
approach to the problem of the color instability by adding extra matter
multiplets to eliminate dangerous contributions  has been proposed in
Ref.~\cite{HKN}.


We begin by  describing the model we are considering. The model is 
similar to that of Dvali and Pomarol \cite{DP} (See also Ref.~\cite{NW}). 
Ref.~\cite{DP} considered a high scale
cut-off theory such as the heterotic string, where 
the Fayet-Illiopoulos (FI) term for the anomalous $U(1)_X$ gauge group 
is given by 
$
   \xi= \frac{{\rm Tr} Q g^2}{192 \pi^2} M_{Pl}^2
$
with $g$ the gauge coupling constant. 
On the contrary, what we will consider is a low-scale cut-off theory
with the cut-off $M_*$. Here we assume that the standard model sector
as well as the $U(1)_X$ is confined on a brane-like object such as a 
D-brane and
the large four dimensional Planck scale requires the existence of 
large extra dimensions in which gravity propagates. 
Then it is natural to expect that the  
 FI term $\xi$ is, if it is non-zero,
\begin{equation}
|\xi| \leq M_*^2.
\end{equation}
{}From now on we will 
take a convention $\xi>0$.  $M_*$
may be identified with the string scale. In type I and type IIB string 
models, 
$\xi$ will be generated through
non-vanishing expectation values of some moduli fields when
combined with a generalized Green-Schwarz anomaly cancellation mechanism
\cite{IRU,P}. 

As for chiral multiplets, we introduce
$\phi_+$ and $\phi_{-}$ with $U(1)_X$ charge $+1$, and $-1$,
respectively, and $y_i$ with charge $Q_i$, which represent fields in
the standard model sector.  Then the $U(1)_X$
D-term is written
\begin{equation}
  D= \xi +|\phi_+|^2-|\phi_-|^2+\sum_i Q_i |y_i|^2
\end{equation}
The model also has the following mass
term in the superpotential
\begin{equation}
   W= m \phi_+ \phi_-,
   \label{massterm}
\end{equation}
besides the superpotential of the standard model sector. Here we
assume $m^2 \leq g^2 \xi$.  Though it is possible to generate the mass
term Eq.(\ref{massterm}) dynamically, we will treat it as a given parameter. 
Note that
this does not mean to introduce a huge hierarchy into mass parameters,
since all the mass scales of this low-scale theory are not very far
from the electroweak scale.

By minimizing the scalar potential of the model
\begin{equation}
  V=\left|\frac{\partial W}{\partial \phi_+} \right|^2
    +\left|\frac{\partial W}{\partial \phi_-} \right|^2
    +\frac{g^2}{2} D^2
\end{equation}
with $g$ being the gauge coupling constant for $U(1)_X$, we find
the following vacuum expectation values 
\begin{eqnarray}
   \phi_+  &=&0, \nonumber \\
   \phi_- &=& \sqrt{\xi-\frac{m^2}{g^2}}, \nonumber \\
   F_{\phi_{+}} &=& m \phi_- = m \sqrt{\xi-\frac{m^2}{g^2}}, \nonumber \\
   F_{\phi_{-}} &=& 0, \nonumber \\
   D  &= & \frac{m^2}{g^2}
\end{eqnarray}
Here we have neglected the contributions from the standard model
sector, which are assumed to be tiny.

In this model the scalar masses in the standard model sector are
written in the following form:
\begin{equation}
   m_0^2=  Q g^2 D + m_F^2 = Q m^2+ m_F^2,
\end{equation}
which are given at the cut-off scale $M_*$. 
The first term is the $U(1)_X$ D-term contribution which is solely
controlled by the $U(1)_X$ charge $Q$. On the other hand, the second
term which represents a F-term contribution comes from non-renormalizable
interaction in the K\"ahler potential and is sensitive to the physics
close to the cut-off scale. In fact we expect to have the following term
in the K\"ahler potential:
\begin{equation}
     \frac{\eta_{ij}}{M_*^2} \phi_+^* \phi_+ y_i^* y_j
\label{eq:non-renormalizable}
\end{equation}
with numerical coefficients $\eta_{ij}$ of order unity or less, which
are in general generation dependent.\footnote{Non-renormalizable
  interactions including bulk fields will be suppressed by the four
  dimensional Planck mass $M_{Pl}$.} Eq.~(\ref{eq:non-renormalizable})
yields a (possibly) generation dependent F-term mass estimated at most
as
\begin{equation}
   m_F^2 \sim \frac{F_{\phi_+}^2}{M_*^2} \sim 
   m^2 \left( \frac{\xi-m^2/g^2}{M_*^2} \right).
\end{equation}
Therefore this is potentially a source for the FCNC. However 
it is always sub-dominant compared to the first term, provided that 
there is a little
hierarchy of one order of magnitude or so between 
$\sqrt{\xi-m^2/g^2}$ and $M_{*}$:
\begin{equation}
      \epsilon \equiv  \frac{\sqrt{\xi-m^2/g^2}}{M_{*}} \leq O(10^{-1}).
\end{equation}

Here it is interesting to mention the mass spectrum of $\phi_+$ and $\phi_-$.
What happens is that both supersymmetry and the $U(1)_X$ gauge symmetry are
broken spontaneously. Since the scalar component of $\phi_-$ acquires 
non-zero vacuum expectation value, its  real component has a similar mass
to the gauge boson mass, $\sqrt{g^2\xi-m^2}$, and its imaginary component
becomes the would-be Nambu-Goldstone boson. On the other hand, 
it is essentially the $\phi_+$
multiplet which is responsible for supersymmetry breaking, and
its spinor component is the Goldstino absorbed into the gravitino in 
supergravity framework. In this case the mass of the scalar component of 
the $\phi_+$ multiplet is found to be $\sqrt{2}m$, similar in size with
the soft supersymmetry breaking masses.


\newcommand{\tm}{\tilde{m}}
\newcommand{\tmsqr}{\tilde{m}^2}
\newcommand{\tmsqrh}{\tilde{m}^2_{1,2}}
\newcommand{\tmh}{\tilde{m}_{1,2}}
\newcommand{\DRbarp}{ {\overline{\textrm{DR}}^{\prime}} }
\newcommand{\MSbar}{\overline{\textrm{MS}}} 
\newcommand{\GeV}{\textrm{GeV}}
\newcommand{\TeV}{\textrm{TeV}}

Let us now discuss how the low cut-off scale model ameliorates the problems 
of the decoupling solution.   First we will consider the color instability. 
For simplicity, we assign $U(1)_X$ charges for quarks and leptons in the 
first-two generations to be $+1$ and those for other matters to be $0$.
In this framework, from the cut-off scale $M_*$ to the mass scale of 
the sfermions in the first two generations $\tilde{m}_{1,2}$ the
nature can be described by four-dimensional field theory 
with the  matter content of the minimal supersymmetric standard model (MSSM),
and from the $\tmh$ scale to about 1 TeV it can be described by an 
effective theory
in which squarks and sleptons in the first-two generations are integrated out.

A constraint on the soft masses in the third generation at the cut-off
scale is obtained by requiring physical masses to be positive at the
electroweak scale.  Although the physical masses receive D-term
contributions of the $SU(2)_L \times U(1)_Y$ gauge interactions and
also left-right mixing effects, we neglect them for simplicity ( the
effects due to these neglect are discussed in \cite{AG} ).  Hereafter
we will obtain a constraint by requiring the running masses to be
positive at 1 TeV scale as was done in \cite{AM,AG}.

The values of soft masses at 1 TeV scale are computed by using
renormalization group equations (RGEs).  In our analysis, we use the
two-loop RGEs in the $\DRbarp$ scheme \cite{2loopRGE}.  For our charge
assignment, the RGEs for the soft masses in the third generation that
include Yukawa couplings, A-terms at one-loop level and heavy sfermion
contributions at  two-loop level in this scheme are
\begin{eqnarray}
  \mu \frac{d \tmsqr_f}{d \mu} &=& -\frac{2}{\pi} \sum_A \alpha_A C^f_A M_A^2 
  + \frac{1}{2 \pi} \eta^t_f \alpha_t 
   ( \tmsqr_{Q_3} + \tmsqr_{U_3} + \tmsqr_{H_u} + A_t^2 )
   \nonumber \\ & &
  + \frac{1}{2 \pi} \eta^b_f \alpha_b 
  ( \tmsqr_{Q_3} + \tmsqr_{D_3} + \tmsqr_{H_d} + A_b^2 )
  + \frac{1}{2 \pi} \eta^{\tau}_f \alpha_b 
   ( \tmsqr_{L_3} + \tmsqr_{E_3} + \tmsqr_{H_d} + A_{\tau}^2 )
   \nonumber \\ & &
 + \frac{2}{\pi^2} \sum_A {\alpha_A}^2 C^f_A \tmsqrh  \quad ,
\label{eq:RGE}
\end{eqnarray}
where $\alpha_A$ and $C_A^f$ are the gauge couplings and the quadratic
Casimir of 
$SU(3)_C$, $SU(2)_L$ and $U(1)_Y$ for $A=3,2,1$, respectively,
and $\alpha_f = Y_f^2/4 \pi (f = t, b, \tau)$ are Yukawa couplings, 
with $\eta^t_f = (1,2,3)$ for $f=Q_3, U_3, H_u$ , 
$\eta^b_f = (1,2,3)$ for $f=Q_3, D_3, H_d$, 
and $\eta^{\tau}_f = (1,2,1)$ for $f=L_3, E_3, H_d$, respectively,
and zero otherwise. Note that the contribution proportional to the $U(1)_Y$
D-term does not appear in  (\ref{eq:RGE}) for the sake of the boundary
conditions for the squark and slepton masses.   
We solve the RGEs as follows.
At the cut-off scale, we assume that gaugino masses satisfy 
\begin{eqnarray}
  M_3(M_*) = \frac{\alpha_3(M_*)}{\alpha_2(M_*)} M_2(M_*)  = 
  \frac{\alpha_3(M_*)}{\alpha_1(M_*)} M_1(M_*) 
\end{eqnarray}
for simplicity. Below the $\tmh$ scale, the heavy sfermions do not contribute 
to the running of the couplings and masses. Note that the gaugino masses 
evolve differently from the gauge coupling constants, but we ignored this 
deviation which is not important in our analysis. 
For the squark sector the gluino contribution
dominate the other gaugino contributions, so that the constraint for
the squark masses is insensitive to this assumption. Because of the
charge assignment mentioned above, the squarks and sleptons in the
first-two generations have soft masses $\tmh \simeq m$ at the cut-off
scale. The soft masses of the third generation scalar bosons and the Higgs 
bosons  have no contribution 
from the $U(1)_X$
D-term and their F-term contributions are model dependent.  Here to
simplify the analysis we assume them all universal: 
$m_0 \simeq \epsilon \tmh \simeq \epsilon m$ at
the cut-off scale. We take into account the bottom and tau Yukawa
couplings as well as the top Yukawa coupling.  In our analysis, we fix
the top quark mass in the $\MSbar$ scheme  
$m_t^{\overline{MS}} (m_t)$ to be 167 GeV 
and that of the bottom quark  
$m_b^{\overline{MS}} (m_b)$ to be 4.3 GeV.  
We checked whether our results depend on $\tan\beta$ 
and  $A$ parameters and found that the
dependence on these parameters are very small.  Thus we will present
results with all $A$ parameters zero at the cut-off scale and $\tan \beta = 2$.

We also have to include finite term contribution 
because the scale at which the initial condition of RGEs is given 
is not much lager 
than the electroweak scale.
We follow Ref.~\cite{PT} to evaluate this effect and 
the result for our case is
\begin{eqnarray}
  \tmsqr_{f,finite} (\mu) = -\frac{1}{\pi^2} ( \frac{\pi^2}{3} - 2 - 
   \ln \left( \frac{\tmsqrh}{\mu^2} \right) ) 
  \sum_A \alpha_A^2  C^f_A \tmsqrh \quad .  \label{eq:finite}
\end{eqnarray}
At the $\tmh$ scale, we add this contribution to the running mass, 
as a threshold effect. The $U(1)_Y$ D-term contribution is absent as 
in the case of Eq.~(\ref{eq:RGE}) .
Note that the finite contribution (\ref{eq:finite}) is different from that of 
Ref.\cite{AG}. The difference can be absorbed by a redefinition of the 
renormalization scale $\mu$. 

In fig. 1, we show the allowed maximum values of the sfermion masses
in the first two generations $\tmh$ by requiring that the mass squared
$\tilde{m}^2_{Q_3}$ be positive at the 1 TeV scale. The horizontal
axis represents the running gluino mass at the scale $\mu = 1$
TeV. Here we take all the scalar masses of the third generation and
the soft masses of the Higgs doublets at the cut-off scale
$\tm_{f}(M_*)$ to be $1 \TeV$ .  We found that the constraints from
the positivity requirements of $\tmsqr_{U_3}$ and $\tmsqr_{D_3}$ are
similar to that from $\tmsqr_{Q_3}$ presented here, and in fact the
differences are less than 10 \%. We consider the cases $M_*=100,$
$10^3$, $10^4$ and $10^5$ TeV. For comparison, we also show the case
$M_*=10^{16} $ GeV. The region above each curve is excluded.  As $M_*$
decreases, one finds that the allowed region becomes larger because
the RGE effect becomes less significant. Indeed the finite term
dominates when $M_*=100$ TeV. When the gluino mass is, for instance,
about 1 TeV, $\tmh$ can be as heavy as about 17 TeV for $M_*=100$ TeV.
In this case the contribution to the kaon mass difference $\Delta m_K$
will become at an acceptable level with the help of a small alignment
between squark mass eigenstates and interaction eigenstates \cite{GGMS}.  The
constraint from the positivity requirement of the third generation
slepton $\tmsqr_{L_3}$ alone is not so strong. For example, $\tmh$ is
required to be smaller than about $80$ TeV and $25$ TeV, for $M_* =
100$ TeV and $M_* = 10^5$ TeV, respectively, as far as the gluino mass 
at the 1 TeV is smaller than 3 TeV.
On the other hand, in the case $M_* = 10^{16}$ GeV $\tmh$ is required 
to be smaller than about $13$ TeV.

Next we would like to discuss how the other difficulties are cured
in our setting.  The point of the naturalness problem discussed 
in Ref.~\cite{DG} is that the heavy first two generation scalar masses will
influence the running of the Higgs mass at two loop level, causing the 
fine tuning  to obtain the electroweak scale if the masses are
very heavy. Now since the contribution to the running is roughly proportional 
to the ``length'' of  running in logarithmic scale, the fine-tuning
problem should be relaxed in our low-scale cut-off case  in which 
the length of running is much shorter than the high-scale cut-off case
discussed by \cite{DG}.  

In our scenario, the gravitino becomes very light with the estimate
\begin{equation}
   m_{3/2} =\frac{F_{\phi_+}}{\sqrt{3}M_{Pl}}
          \simeq m \frac{\sqrt{\xi-m^2/g^2}}{M_{Pl}}
          \simeq 0.1 {\rm eV} \left( \frac{M_*}{100 {\rm TeV}} \right)
                            \left( \frac{\epsilon m}{1 {\rm TeV}} \right).
\end{equation} 
Thus it is likely to be the lightest superparticle (LSP). Then the lightest 
superpartner in the supersymmetric standard model sector is no longer stable,
but it immediately decays into the gravitino and hence it is obvious that 
the overclosure problem of the neutralinos is evaded. 
In this scenario, the gravitino is stable. Its cosmological implications
are discussed in the literature. 
For the gravitino which weighs much less than 1 keV, its relic 
abundance is much smaller than the critical density of the Universe and
thus it is cosmologically harmless.
See Ref. \cite{grav-cosmology} and references therein for detail.
We should also note that superparticle signals at collider
experiments in our scenario
have some characteristic features. The lifetime of the next to the lightest
superparticle (NSP) is roughly of the order
\begin{equation}
  \tau_{\rm NSP} \simeq
    16 \pi \frac{F_{\phi_+}^2}{m_{\rm NSP}^5} \simeq 
    10^{-17} {\rm sec}
     \left( \frac{100 \GeV}{m_{\rm NSP}} \right)^5
     \left( \frac{\epsilon m}{1 \TeV} \right)^2
     \left( \frac{M_*}{100 \TeV} \right)^2 ,
\end{equation}
assuming that the decay is a two-body decay.
Thus the lifetime is so short  that it will decay inside a detector. If
the NSP is bino-like, the decay contains a photon and a gravitino which escapes
detection. If the NSP is a slepton, most likely a stau, the decay contains
a tau lepton and a gravitino. In our scenario, the stop may be the NSP. In 
this case, the
stop decays into a top (or a W boson and a bottom quark\footnote{For the 
three-body decay, the decay length increases substantially. 
An analysis in this case has been given in  \cite{CP}.}) and a gravitino.
In either case the signals will be
distinguishable from those of high-scale supersymmetry breaking scenario where
the signals will be associated with a massive LSP escaped from a detector.
Here it should also be noticed that we can probe the heavy mass scale of the 
first-two generations via superoblique corrections \cite{superoblique}, 
even though they cannot be produced directly in near future colliders.

Here  we would like to briefly mention a
gaugino mass. In our model we can write the following term 
\begin{equation}
   \frac{\phi_+ \phi_-}{M_*^2} W^{\alpha}W_{\alpha}, 
\end{equation}
where $W^{\alpha}$ is a supersymmetric field strength of a gauge field.
It follows from this that the gaugino mass is of the order
\begin{equation}
     \frac{F_{\phi_+} \phi_-}{M_*^2}= \epsilon^2 m.
\label{eq:gaugino}
\end{equation}
Recall that the mass of the third generation squark is $\sim \epsilon m$. 
An additional suppression factor $\epsilon$ in Eq.~(\ref{eq:gaugino})
may be compensated by unknown numerical coefficients in front.
Note that in the low-scale supersymmetry
breaking scenario, the contribution to the gaugino mass
from superconformal anomaly \cite{gauge-mediation}
is negligible because it is proportional to
the tiny gravitino mass. 

Before concluding we will comment on the large extra dimensions needed
to obtain the large Planck scale with the low string scale
scenario.  The size of the compact $n$-dimensional extra dimensions $R$
is given by
\begin{equation}
        M_{Pl}^2 \simeq M_*^{2+n} R^n,
\end{equation}
or
\begin{equation}
       R^{-1} \simeq M_* \left(\frac{M_*}{M_{Pl}}\right)^{2/n}
\end{equation}
to reproduce the Planck scale $M_{Pl}$. Here we have assumed that 
the compact manifold is isotropic and is characterized by a single size $R$.
To illustrate, let us take $n=6$. Then
\begin{equation}
          R^{-1} \simeq 10^2 \mbox{GeV} 
                 \left( \frac{M_*}{10^6 \mbox{GeV}} \right)^{4/3}.
\end{equation}
The masses of graviton's Kaluza-Klein modes are quantized in units of 
$R^{-1}$. Thus we find that the KK mode masses are in the electroweak scale
or higher.  This contrasts with the case of the large extra dimension
scenario with $M_* \simeq 1$ TeV where $R^{-1}\simeq 10$ MeV.

Since the KK modes have masses of the electroweak scale or so and have
interactions similar to the graviton, they may affect cosmological
evolution of the early Universe.  In particular, they are produced after
the inflationary epoch and decay typically around the
epoch of the big-bang nucleosynthesis. Here we will not go into
detailed discussion, but make some remarks.  First if $R^{-1} > 10^4$ GeV,
the KK modes decay before the nucleosynthesis and
thus they are harmless. On the other hand, for $10^2 \mbox{GeV} < R^{-1} 
<10^4$ GeV, the reheat temperature after inflation must be low to suppress the 
production of the KK modes. We expect that the reheat temperature of $10^2$ 
GeV will be allowed since then the production of the KK modes whose masses 
are heavier than $10^2$ GeV is highly suppressed. This is very different from
the TeV gravity case where the reheat temperature is forced to be even
smaller than 1 GeV \cite{ADD}. The higher reheat temperature in our case has an
advantage for baryogenesis. In particular one may be able to use 
the electroweak baryogenesis.  

It is interesting to mention here that the radius of the extra dimension
can be as large as a sub-millimeter for $n=1$ and $M_* \simeq 10^8$ GeV.
This case may be tested in a future gravity experiment \cite{ADD}.

There remains a problem of how one realizes a viable inflation model and a 
subsequent  stabilization of the size of the extra dimensions. A hope is
that model building for this may be somewhat easier than the original
large extra dimension scenario \cite{ADKM}. This issue should
deserve further study. 

To summarize, we have considered the supersymmetric standard model
with the anomalous $U(1)$ gauge symmetry when the ultraviolet cut-off
scale is not far from the electroweak scale.  In our scenario, the
Fayet-Illiopoulos D-term is set to be a bit smaller than the cut-off
scale squared. Except this, the model is similar to that of \cite{DP}.
We applied this model to the decoupling solution of the supersymmetric
flavor problem and showed that the difficulties of the solution become
less severe than the conventional high-scale cut-off scenario.  The
model should be combined with the idea of the large extra dimensions
to obtain the large Planck scale of the gravitational interaction. We
briefly discussed some of the related cosmological issues. 

We would like to thank K. Kurosawa and Y. Nomura for valuable
discussions and explaining the works in \cite{HKN}. MY also thanks
Y. Kawamura and H. Nakano for helpful discussions on anomalous U(1)
theories in type I and type IIB string models. The work of YY was
supported in part by the Grant-in-Aid for Scientific Research from the
Ministry of Education, Science, Sports, and Culture of Japan,
No.10740106, and the work of MY by the Grant-in-Aid on Priority Area
707 "Supersymmetry and Unified Theory of Elementary Particles", and by
the Grant-in-Aid No.11640246.

\begin{figure}[b]
  \begin{center}
    \includegraphics[height=8cm,clip]{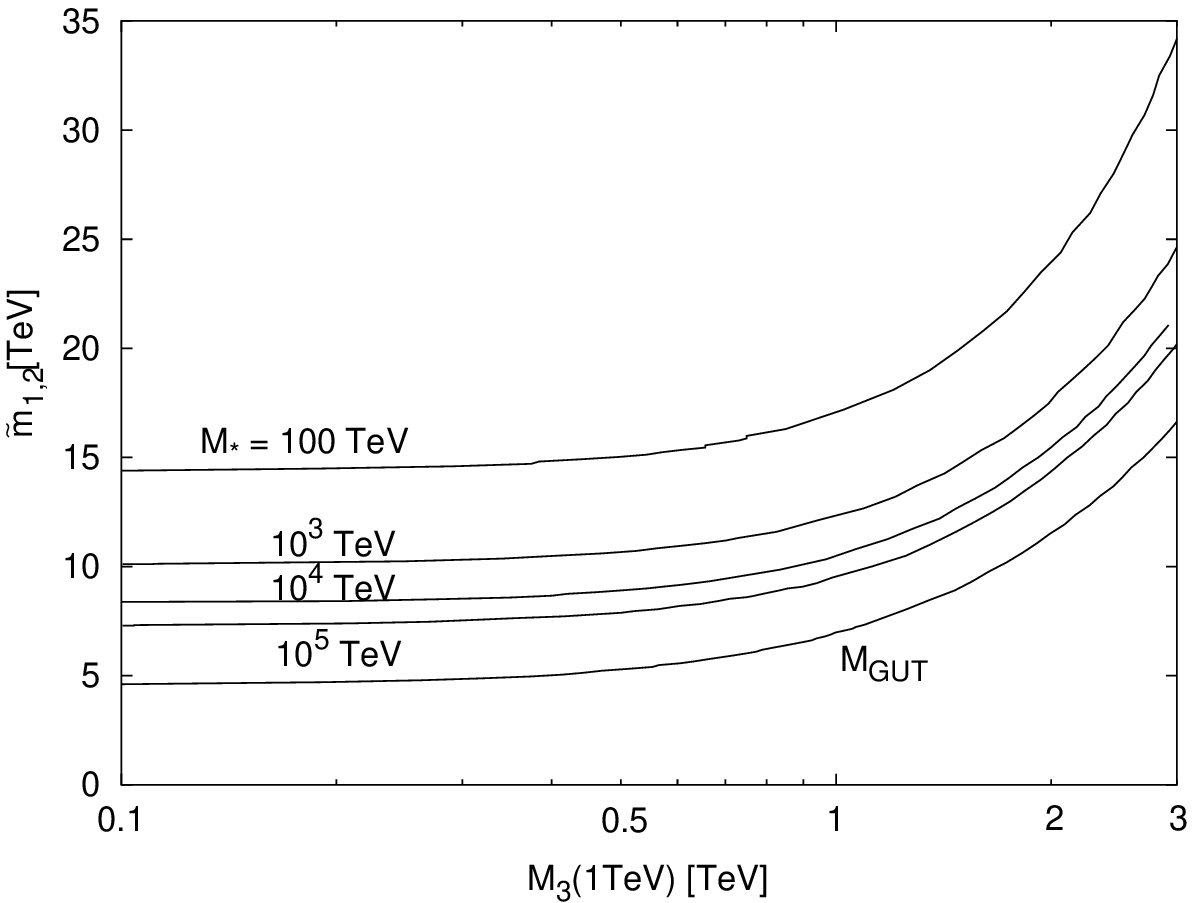}
    \vspace{1 cm}
    \caption{The allowed maximum values of the sfermion masses in the 
      first two generations $\tmh$ by requiring that the stop mass squared
      $\tilde{m}^2_{Q_3}$ be positive at the 1 TeV scale, for the cut-off 
      scale $M_* = 100, 10^3, 10^4, 10^5$ TeV and 
      $M_* = M_{\rm{GUT}} = 10^{16}$ GeV. 
      The region above each curve is excluded. The horizontal
      axis represents the running gluino mass at the scale 1 TeV. 
      All the scalar masses of the third generation as well as  
      the soft masses of the Higgs doublets at the cut-off scale are 
      fixed to be 1 TeV. }
  \end{center}
\end{figure}

\end{document}